\documentclass{article}
\usepackage[utf8]{inputenc}
\usepackage[T1]{fontenc}
\let\bwlOriginalAddcontentsline\addcontentsline
\PassOptionsToPackage{numbers,sort&compress}{natbib}
\usepackage[nusbwlab]{iclr2025_conference}
\let\addcontentsline\bwlOriginalAddcontentsline
\usepackage{times}
\usepackage{amsmath,amssymb,amsfonts,amsthm,mathrsfs}
\usepackage{textcomp,booktabs,multirow,nicefrac,microtype}
\usepackage{xcolor,graphicx,url}
\usepackage{algorithm,algpseudocode,float,placeins}
\usepackage{needspace}
\setcitestyle{numbers,square,citesep={,}}
\usepackage[hidelinks]{hyperref}
\graphicspath{{figures/}{figs/}}
\renewcommand{\bibnumfmt}[1]{#1.}
\date{}
\title{STAMP as a topology-aware molecular language for fragment-level learning and editing}
\author{\normalsize Yuxuan Ren\textsuperscript{1}, Jiatong Li\textsuperscript{1}, Ziqiao Meng\textsuperscript{1}, Yatao Bian\textsuperscript{1,$\dagger$}\\[0.6em]\normalsize\textsuperscript{1}National University of Singapore\\[0.3em]\normalsize\textsuperscript{$\dagger$}Correspondence to: \href{mailto:ybian@nus.edu.sg}{ybian@nus.edu.sg}}

\begin{document}
\maketitle
\begin{abstract}
Molecular fragments provide chemically meaningful units, but existing molecular languages rarely make their locations and connections explicit for editing. Here we introduce STAMP (Sequential Tree Attachment-based Motif-Preserving representation), a topology-aware molecular language that connects fragment-level representation learning with precise structural editing. STAMP separately encodes motif identity, attachment sites and tree topology, making fragments reusable across contexts and subtrees directly accessible. Learned representations capture motif similarity and recurring transformations, and lower activity-cliff prediction error on 23 of 30 targets versus a matched SMILES encoder. Compatible subtree operations support scaffold-preserving optimization, ranking first among six methods on 15 of 23 benchmark objectives. A deterministic compiler enables frozen language models to act on indexed fragments, improving molecular editing, optimization, transformation transfer and change identification by 9 to 56 percentage points over plain SMILES. Together, these results establish an executable molecular language in which learned chemical units remain directly operable during design.
\end{abstract}

\section{Introduction}

Language models trained on molecular strings can generate compounds \cite{segler2018rnn,bagal2022molgpt}, predict molecular properties \cite{ross2022molformer,ahmad2022chemberta2} and assist synthesis planning \cite{segler2018synthesis,schwaller2019moltransformer}. The molecular language determines which chemical units and relations are explicit during learning. Reliable reconstruction and validity are fundamental requirements \cite{oboyle2018deepsmiles}. Molecular design also requires locating and modifying chemical units while preserving the surrounding structure. SMILES faithfully serializes molecular graphs, but a functional-group replacement can require coordinated changes across non-contiguous parts of the string \cite{weininger1988smiles}. SELFIES addresses validity \cite{krenn2020selfies,skinnider2024invalid}; fragment-level editing additionally requires explicit attachment sites and substructure boundaries.

Fragment representations expose larger chemical units for learning and construction \cite{erlanson2016fragments,jin2018jtvae,jin2020hiervae,maziarz2022moler}. SAFE, fragSMILES, Group-SELFIES and Molcano make fragments or their connections explicit \cite{noutahi2024safe,fragsmiles2024,groupselfies2023,na2026molcano}. Junction-tree models encode fragment topology \cite{jin2018jtvae,jin2020hiervae}; t-SMILES serializes SMILES-described fragments through a breadth-first traversal of a full binary tree \cite{wu2024tsmiles}. Subsequent work has examined how traversal and fragment order affect generative learning \cite{wu2024tsis}. The question is how to make these chemical units directly addressable for editing while retaining them as units for molecular learning.

An edit must identify the part to change, specify its replacement and preserve the surrounding structure. Fragment identity, attachment sites and assembly topology therefore have distinct roles. Many current serializations couple some of these variables or leave fragment positions implicit \cite{na2026molcano,wu2024tsmiles,wu2024tsis,kong2022psvae,geng2023micam}. Replacement, linker insertion and scaffold modification often depend on external graph operations or regeneration of the molecular string. Representing these variables separately can make the target, boundary and attachment requirements of an edit explicit within the sequence.

Here we introduce STAMP, the Sequential Tree Attachment-based Motif-Preserving representation, as an ordered fragment-tree molecular state. Distinct motif, joint and branch tokens encode fragment identity, attachment atoms and tree topology. A benzene motif can therefore be reused across attachment positions and substitution contexts. A canonical depth-first traversal assigns fragment indices and maps each subtree to a contiguous token span. Deterministic decoding assembles the encoded fragments into a molecular graph, while the indexed fragments and subtree spans define targets for structural editing (Fig.~\ref{fig:stamp_overview}).

The same state supports both molecular learning and structural operations. Generative and predictive models learn from its motif identities and structural relations. Search algorithms replace an indexed subtree with an attachment-compatible donor and decode the resulting sequence. For instruction-guided editing, a frozen language model selects an indexed target and a growth, deletion or replacement action. A deterministic compiler checks compatibility, executes the action and validates the product. The chemical units exposed during learning thus remain accessible to search algorithms and language models during molecular modification.

We evaluate this shared state through generation, property prediction, optimization and instruction-guided editing. STAMP exactly reconstructs all 20,000 molecules sampled from the ZINC-250K training split and produces compact sequences. Under matched training and sampling, it gives the highest yield of unique valid molecules meeting drug-likeness and synthetic-accessibility thresholds among eight evaluated representations. Larger-scale, fragment-constrained and text-conditioned experiments extend these construction settings. STAMP encoders improve on ChemBERTa-2 10M baselines on most evaluated MoleculeNet tasks \cite{ahmad2022chemberta2,wu2018moleculenet} and lower activity-cliff prediction error on 23 of 30 MoleculeACE targets relative to the matched SMILES-BPE encoder \cite{vantilborg2022moleculeace}. Their embeddings capture motif similarity, distinguish attachment changes from motif substitutions and represent recurring transformations consistently across molecular contexts.

Fragment-tree search achieves aggregate PMO performance comparable to GenMol and gives the better outcome in 24 of 27 lead-optimization cells where either method succeeds. Compiled actions improve single-step editing with frozen language models and support transfer of demonstrated transformations to new molecules. Together, these results establish STAMP as a common representation through which molecular structure can be generated, learned and modified.

\section{Results}

\subsection{Addressable fragment-tree representations}

\begin{figure}[tbp]
\centering
\includegraphics[width=\linewidth,height=0.68\textheight,keepaspectratio]{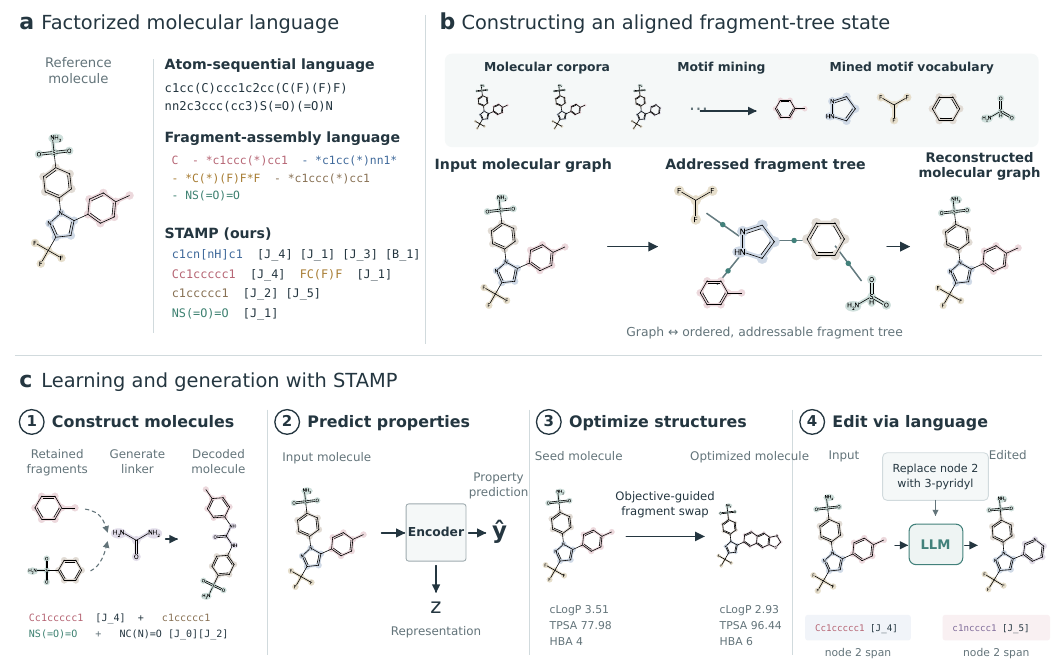}
\caption{\textbf{A factorized fragment-tree language connects molecular structure, sequence, learning and action.} \textbf{a}, The reference molecule in an atom-sequential language, a fragment-assembly language and STAMP. Motif tokens encode fragment identity; joint and branch tokens encode attachment sites and tree topology. \textbf{b}, Motif mining, encoding into an addressed fragment tree and deterministic reconstruction. Motif groups align tree nodes with sequence spans. \textbf{c}, Four applications of the shared state: fragment-constrained molecular construction, property prediction, objective-guided fragment optimization and instruction-guided editing. For editing, the language model selects a node and fragment, and the compiler executes the action.}
\label{fig:stamp_overview}
\end{figure}

STAMP represents a molecule as a rooted fragment tree with distinct motif, joint and branch tokens (Fig.~\ref{fig:stamp_overview}a,b). Motif tokens encode canonical fragments, joint tokens identify the atoms involved in cross-fragment bonds, and branch tokens specify the ordered tree topology. Separating attachment sites from fragment identity allows the same motif token to serve multiple substitution contexts.

Each motif group and each descendant subtree occupies a contiguous token span. Node indices and attachment ports locate an edit, while bond order and open-port count define replacement compatibility. These addressable units support molecular construction, property learning, goal-directed optimization and instruction-guided editing (Fig.~\ref{fig:stamp_overview}c).

The three token classes specify complementary aspects of molecular assembly. A motif identifies a chemical fragment, a joint selects an atom at its boundary, and a branch delimits the descendants connected through that boundary. Changing an attachment site can therefore preserve motif identity, while replacing a subtree changes a defined fragment group together with its descendants. Within each encoded molecule, a node index points to the selected motif and the subtree span identifies the sequence region affected by the operation. Tree construction and stack-machine decoding provide the mapping between these addresses and the molecular graph (Supplementary Note~1).

STAMP exactly reconstructed all 20,000 molecules (100\%) in the common sample from the ZINC-250K training split. On this sample, it had the shortest mean sequence among eight representations: 15.31 tokens, with a 95th percentile of 24 (Supplementary Table~1). SMILES, SELFIES and Group-SELFIES also reconstructed every sampled molecule exactly; their mean sequences were more than twice as long. fragSMILES averaged 16.66 tokens with 93.345\% exact reconstruction; SAFE reached 96.570\%. t-SMILES TSDY and Molcano achieved valid-decoding rates of 99.970\% and 100\%, respectively, with stereochemistry-sensitive exact-reconstruction rates of 15.925\% and 57.085\%.

Across 1,497 ChEMBL matched molecular pairs, STAMP edits affected an average of 2.24 contiguous spans, compared with 2.31 for fragSMILES and 4.07 for SMILES. The largest changed span averaged 5.60, 6.28 and 10.10 tokens, respectively (Supplementary Fig.~2). STAMP localized matched edits to fewer and shorter spans than SMILES, with locality comparable to fragSMILES.

\subsection{Molecular generation across conditioning settings}

We evaluated STAMP for de novo generation, fragment-constrained construction and text-to-molecule generation.

\begin{figure}[tbp]
\centering
\includegraphics[width=\linewidth,height=0.68\textheight,keepaspectratio]{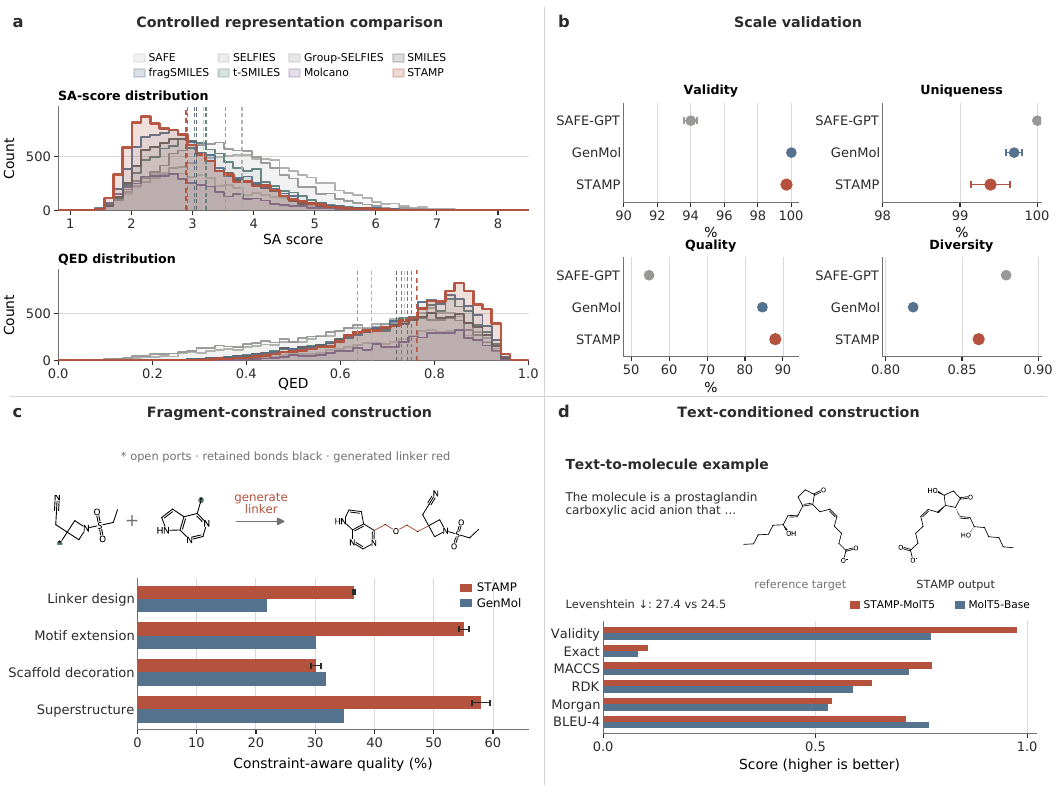}
\caption{\textbf{STAMP improves molecular generation across unconditional and conditioned settings.} \textbf{a}, SA and QED histograms for eight representations trained with the same four-layer causal Transformer (dimension 256) for 10 epochs with a shared 100,000-molecule ZINC-250K training input set. Each model produced 10,000 requested samples. Histograms count unique valid molecules in common bins; dashed lines indicate means. \textbf{b}, Larger-scale generation with a 12-layer STAMP AO-GPT (dimension 768). Points show mean \ensuremath{\pm} s.d., using five seeds for STAMP and published values for SAFE-GPT and GenMol. The displayed GenMol configuration uses \ensuremath{\tau} = 0.5 and r = 0.5. \textbf{c}, Constraint-aware quality for four fragment-constrained tasks. The linker example retains the supplied fragments and ports and generates the highlighted connecting path. STAMP error bars show s.d. across five seeds; GenMol values are published means. \textbf{d}, A ChEBI-20 text-to-molecule example and six aggregate metrics on a common 0--1 scale. Levenshtein distance is shown separately. The reference molecule is the evaluation target.}
\label{fig:molecular_construction}
\end{figure}

We compared eight representations using matched four-layer causal Transformer backbones trained on the same 100,000 ZINC-250K inputs. Each model received 10,000 generation requests under a shared sampling and post-processing protocol. Quality yield was the fraction of requests producing a unique valid molecule satisfying QED \ensuremath{\geq} 0.6 \cite{bickerton2012qed} and SA \ensuremath{\leq} 4 \cite{ertl2009sascore}.

STAMP achieved the highest quality yield, 75.5\%, compared with 66.8\% for fragSMILES, 58.4\% for t-SMILES TSDY, 55.2\% for SMILES and 31.4\% for Molcano (Fig.~\ref{fig:molecular_construction}a). Its validity was 97.38\%, whereas SELFIES and Group-SELFIES achieved 100\%. STAMP therefore converted more of the same generation budget into distinct molecules meeting both property thresholds. Supplementary Table~2 gives the complete single-run results.

For larger-scale generation, we trained a 12-layer any-order autoregressive Transformer (AO-GPT) on STAMP sequences \cite{xue2025aogpt}. Under GenMol's evaluation protocol, STAMP achieved quality of 88.0 \ensuremath{\pm} 1.1\% and diversity of 0.861 \ensuremath{\pm} 0.001, compared with 84.6 \ensuremath{\pm} 0.8\% and 0.818 \ensuremath{\pm} 0.001 for the highest-quality reported GenMol configuration (Fig.~\ref{fig:molecular_construction}b and Supplementary Table~3). STAMP validity was 99.7 \ensuremath{\pm} 0.2\% and uniqueness was 99.4 \ensuremath{\pm} 0.3\%, both within 0.3 percentage points of GenMol.

The AO-GPT formulation also supports completion around retained molecular context. We adapted separate checkpoints to group-aligned infilling: retained motif groups specify the supplied structure, and holes mark the sequence spans to generate. Because each hole follows motif-group boundaries, its completion can be inserted into the existing tree and decoded with the retained context. This gives fragment constraints a direct representation in the model input. The vocabulary and training configurations for the generation and infilling models are summarized in Supplementary Note~5.

We used GenMol's linker-design, scaffold-morphing, motif-extension, scaffold-decoration and superstructure tasks \cite{lee2024genmol}. Linker design and scaffold morphing share the same formulation, giving four distinct settings. Retained fragments formed the fixed STAMP context, and group-aligned holes specified the regions to generate. Constraint-aware quality required a unique valid molecule to preserve the supplied constraints and satisfy the QED and SA thresholds (Fig.~\ref{fig:molecular_construction}c).

STAMP achieved higher constraint-aware quality than GenMol for superstructure generation (58.0 versus 34.8\%), motif extension (55.1 versus 30.1\%) and linker design (36.6 versus 21.9\%). Scaffold-decoration quality was 30.1\% for STAMP and 31.8\% for GenMol. Uniqueness was higher in all four settings, with validity at or above 98.2\% (Supplementary Fig.~1 and Supplementary Note~2). The largest quality gains occurred in tasks that add or replace complete fragments, matching the units exposed by STAMP.

We replaced the molecular strings in the MolT5 recipe with STAMP sequences, retaining the C4+ZINC pretraining mixture, Base-scale architecture and ChEBI-20 evaluation protocol \cite{edwards2022molt5}. Published T5-Base and MolT5-Base results provided the reference baselines.

STAMP-MolT5 outperformed MolT5-Base on five of seven text-to-molecule metrics (Fig.~\ref{fig:molecular_construction}d). Molecular validity increased from 77.2\% to 97.6\%, and canonical-SMILES exact match increased from 8.1\% to 10.6\%. All three fingerprint similarities also increased (Fig.~\ref{fig:molecular_construction}d). The gains thus covered valid molecular construction, exact identity and structural similarity.

\subsection{Property prediction and multiscale chemical organization}

We evaluated STAMP encoders on eight MoleculeNet tasks \cite{wu2018moleculenet} and the 30-target MoleculeACE activity-cliff benchmark \cite{vantilborg2022moleculeace}, then examined their frozen embeddings at the motif, attachment, transformation and whole-molecule levels.

\begin{figure}[tbp]
\centering
\includegraphics[width=\linewidth,height=0.68\textheight,keepaspectratio]{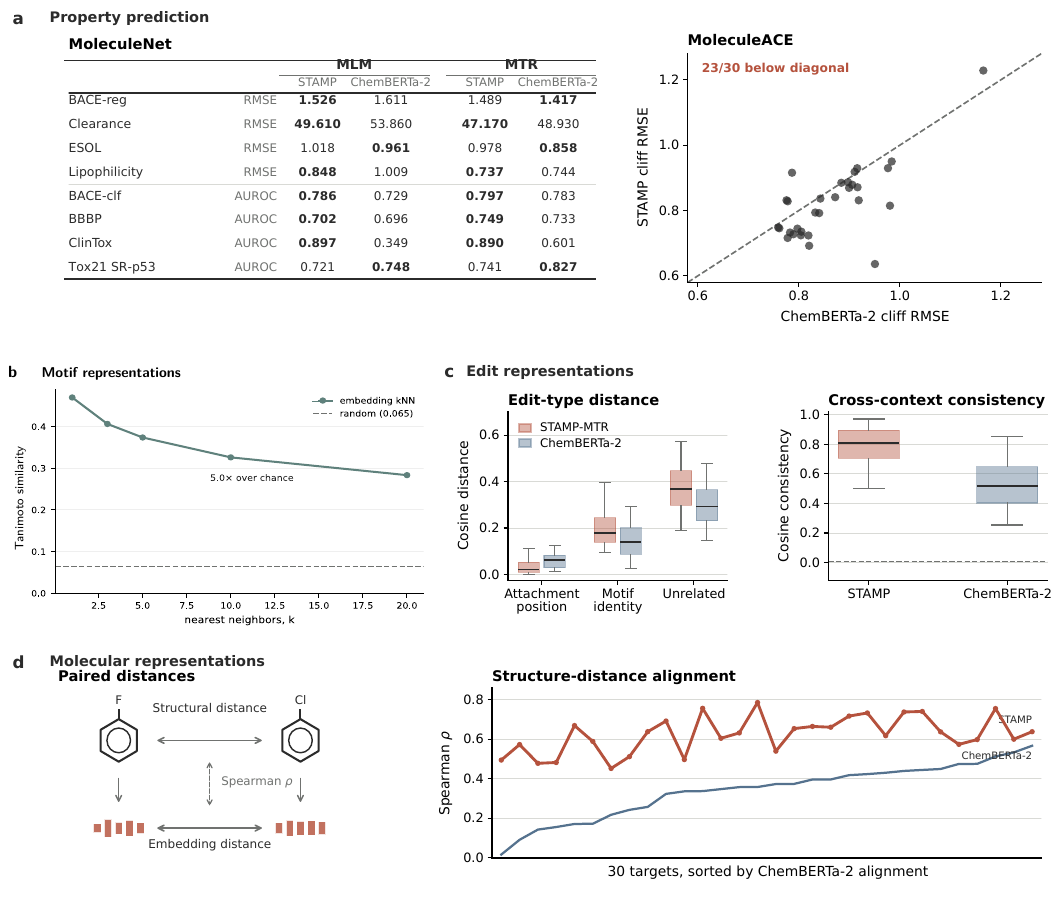}
\caption{\textbf{STAMP improves property prediction and organizes chemical structure at multiple scales.} \textbf{a}, MoleculeNet results for STAMP and published ChemBERTa-2 10M baselines, paired by MLM or MTR objective; bold indicates the better value. The scatter plot compares activity-cliff RMSE on 30 MoleculeACE targets after matched fine-tuning. \textbf{b}, Mean ECFP4 Tanimoto similarity of contextual motif neighbours as neighbourhood size increases, with random motif pairs as the reference. \textbf{c}, Frozen-embedding distances for attachment-position, motif-identity and unrelated molecular changes, followed by the leave-one-context-out consistency of recurring transformations. Boxes span the 25th--75th percentiles, centre lines mark medians, and whiskers extend to the most extreme observations within the 5th--95th percentile interval; outliers are not shown. The dashed null uses displacements from different transformations. \textbf{d}, Structural and embedding distances used in the alignment analysis, and target-level Spearman correlations on the structure-matched subset of 90,620 pairs.}
\label{fig:property_representation}
\end{figure}

With matched encoder configurations and pretraining objectives, STAMP outperformed the published ChemBERTa-2 10M baseline on six of eight MoleculeNet tasks after masked language modelling (MLM) and five of eight after multi-task regression (MTR) (Fig.~\ref{fig:property_representation}a). Supplementary Note~6 and Table~6 give all task-level RMSE and ROC-AUC values.

On MoleculeACE, STAMP-MTR achieved lower full-set and activity-cliff RMSE on 23 of 30 targets than the matched, released SMILES-BPE encoder. The mean paired difference in activity-cliff RMSE was -0.040 (STAMP minus ChemBERTa-2; Wilcoxon p = 0.0040). Across targets, STAMP-MTR obtained mean full-set and activity-cliff RMSEs of 0.719 and 0.825, the lowest among the evaluated deep-learning methods. SVM--ECFP4 \cite{yang2019dmpnn,rogers2010ecfp} was the strongest classical reference, with corresponding RMSEs of 0.671 and 0.751. Supplementary Note~7 and Table~7 place these results alongside the full set of classical and deep-learning baselines.

Contextual STAMP-MTR embeddings grouped chemically related motifs. At a neighbourhood size of ten, the mean ECFP4 Tanimoto similarity among embedding neighbours was approximately five times that of random motif pairs (Fig.~\ref{fig:property_representation}b). Representative neighbourhoods shared ring size, aromaticity and heteroatom patterns across contexts (Supplementary Fig.~4).

We compared frozen embeddings from STAMP-MTR and the matched ChemBERTa-2 PubChem-10M SMILES-BPE encoder on controlled molecular pairs. Attachment-position pairs changed the attachment atom while fixing the parent and probe motifs; motif-identity pairs changed the probe while fixing the parent and attachment atom. Across 207 attachment-position and 1,080 motif-identity pairs, median STAMP-MTR cosine distances were 0.021 and 0.179, respectively, compared with 0.061 and 0.138 for ChemBERTa-2 (Fig.~\ref{fig:property_representation}c). Relative to motif substitutions, attachment changes produced smaller embedding distances in STAMP-MTR: the positional-to-identity ratio was 0.119, versus 0.439 for ChemBERTa-2. The controlled pairs distinguish changing where a motif connects from changing which motif is present. The smaller ratio for STAMP-MTR indicates that its learned geometry preserves this distinction: alternative attachment positions remain closer than substitutions of fragment identity. This organization follows the two structural variables represented by joint and motif tokens.

Repeated transformations produced consistent embedding displacements across molecular contexts. For each transformation, we compared its normalized displacement in one context with the centroid from the other contexts. Across 1,485 transformations observed in at least six contexts, the median leave-one-context-out cosine similarity was 0.808 for STAMP-MTR and 0.520 for ChemBERTa-2. Displacements from different transformations had a median similarity of 0.006 for both encoders. STAMP-MTR thus represented recurring transformations more consistently across contexts while distinguishing unrelated transformations. This analysis examines how embeddings move under an edit. A shared displacement direction links the same fragment transformation across different molecular surroundings, complementing the proximity of chemically related motifs. Agreement with the remaining-context centroid shows that this relationship recurs across the observed contexts of a transformation.

Across 314,873 MoleculeACE matched molecular pairs with valid embeddings from both encoders, the Spearman correlation between embedding cosine distance and ECFP4 Tanimoto distance was 0.671 for STAMP-MTR and 0.386 for ChemBERTa-2. On a structure-matched subset of 90,620 pairs, STAMP retained the higher correlation in all 30 target-level comparisons (Fig.~\ref{fig:property_representation}d). Together, these analyses show that STAMP embeddings organize chemical structure at the motif, attachment, transformation and whole-molecule scales. Supplementary Fig.~3 presents the structure-matched activity-cliff analysis, and Supplementary Note~10 and Table~8 report the RePRA comparison.

\subsection{Executable fragment-tree search for molecular optimization}

We next tested the fragment tree as a search space, first measuring the validity and scaffold retention of subtree replacement, then evaluating molecular optimization on the 23-objective Practical Molecular Optimization (PMO) benchmark and GenMol's docking-based lead-optimization benchmark \cite{lee2024genmol,gao2022pmo}.

\begin{figure}[tbp]
\centering
\includegraphics[width=\linewidth,height=0.68\textheight,keepaspectratio]{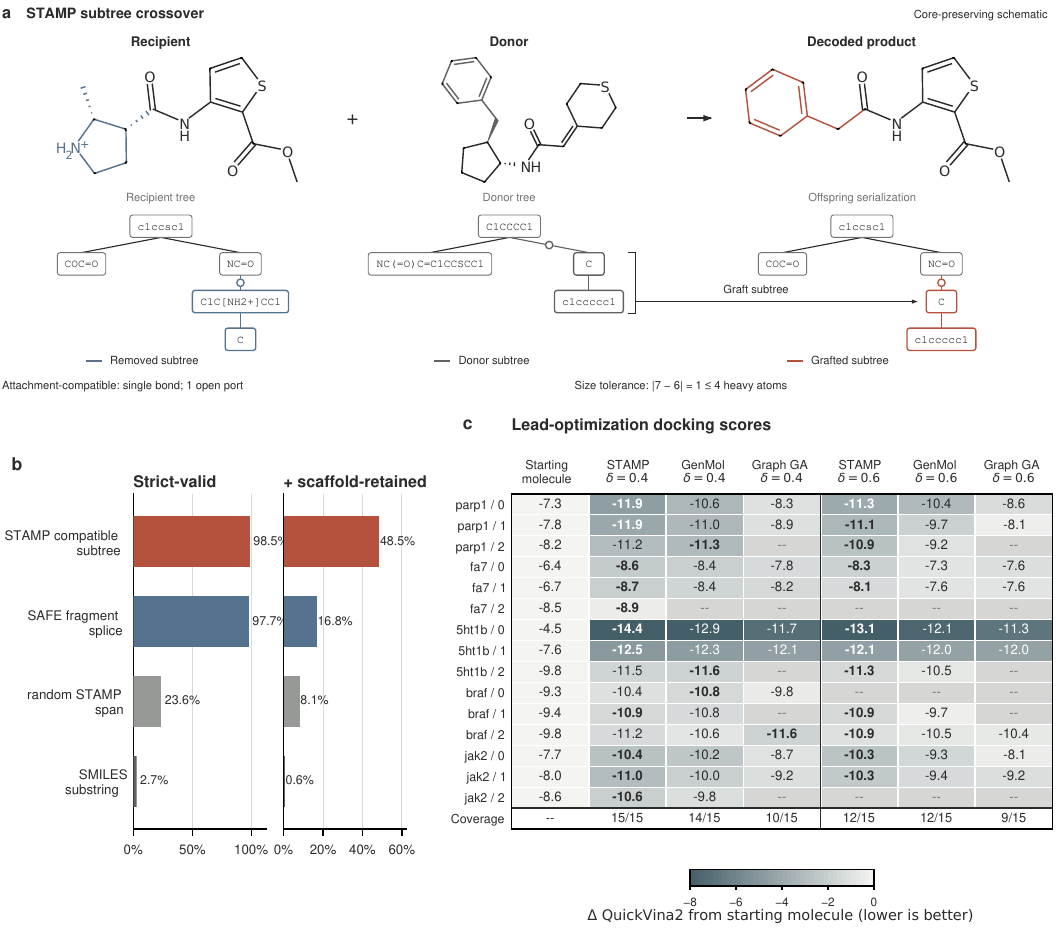}
\caption{\textbf{Fragment-tree search supports executable, core-constrained molecular optimization.} \textbf{a}, STAMP subtree crossover under the default exocyclic-single-bond decomposition. Each box is a motif token and each edge is a parent--child relation. The selected recipient subtree (blue) and compatible donor subtree (dark grey) are contiguous two-node spans. Open circles mark the matched single-bond attachments; heavy-atom size tolerance is checked separately. The right-hand tree shows the offspring serialization immediately after crossover; deterministic decoding yields the molecular product above. \textbf{b}, Four replacement operators evaluated over 1,000 recipient--donor trials each. Both strict validity and strict-valid scaffold retention use all trials as the denominator. \textbf{c}, QuickVina2 scores on the lead-optimization benchmark. Rows identify the target and starting molecule. Cell text gives the absolute score; colour gives the change from the starting molecule. Lower scores are better, and bold marks the best score within each three-method group at similarity threshold \ensuremath{\delta} = 0.4 or 0.6. Dashes indicate no improving candidate satisfying all constraints. Coverage is reported separately. STAMP values are three-seed means; baseline values are from GenMol. The 5ht1b/2 result at \ensuremath{\delta} = 0.6 was reached by the atom-level action.}
\label{fig:optimization}
\end{figure}

TreeCrossover replaces a non-root recipient subtree with a donor subtree having the same incoming bond order and open-port count, subject to a heavy-atom size tolerance (Fig.~\ref{fig:optimization}a). The subtree span localizes the replacement, and the attachment signature determines how it joins the retained structure.

Across 1,000 recipient--donor trials per operator, compatible STAMP replacement produced strict-valid offspring without decoder fallback in 98.5\% of trials. SAFE fragment splicing achieved 97.7\%, arbitrary STAMP-span replacement 23.6\% and SMILES-substring replacement 2.7\% (Fig.~\ref{fig:optimization}b). Using all trials as the denominator, 48.5\% of STAMP replacements were both strict-valid and retained the recipient Bemis--Murcko scaffold, compared with 16.8\% for SAFE, 8.1\% for arbitrary STAMP spans and 0.6\% for SMILES substrings. Among scaffold-evaluable strict-valid products, scaffold retention was 49.2\% for compatible STAMP replacement, 17.3\% for SAFE, 34.5\% for arbitrary STAMP spans and 22.2\% for SMILES substrings. The advantage therefore extended to preservation of the recipient scaffold among successfully decoded products. Together, the two measures show how selecting an explicit subtree boundary improves the yield of useful structural proposals.

We then used these proposals in an iterative search. At each generation, a scored population supplies recipient molecules, and compatible fragments provide local alternatives at selected subtrees. High-reward molecules contribute their subtrees to a donor pool indexed by attachment signature. Subsequent proposals can thus reuse fragments discovered earlier in the search while retaining the recipient context. The optimization loop and donor-pool update are given in Supplementary Note~3.

PMO evaluates 23 black-box objectives with a 10,000-call online oracle budget. Its AUC-Top10 metric captures both the quality of the best molecules and how early they are found. Under the same target-aware prescreening protocol as GenMol, STAMP achieved an aggregate AUC-Top10 of 18.668, compared with 18.362 for GenMol and 16.928 for f-RAG (Table~\ref{tab:pmo_main}). STAMP achieved the highest displayed mean on 15 objectives in the six-method comparison and exceeded GenMol on 17 objectives.

\begin{table}[tbp]
\centering
\caption{\textbf{AUC-Top10 on the Practical Molecular Optimization benchmark.} STAMP values are means across three seeds under a 10,000-call online oracle budget. Bold marks the highest mean in the six-method comparison. STAMP uses the same target-aware prescreening protocol as GenMol; offline-bank construction is specified in Methods. Baseline means and aggregate scores are taken directly from the original publications \cite{lee2024genmol,gao2022pmo,frag2024}.}
\label{tab:pmo_main}
\scriptsize
\setlength{\tabcolsep}{3.3pt}
\renewcommand{\arraystretch}{0.92}
\begin{tabular}{l|cccccc}
\toprule
Oracle & \textbf{STAMP} & GenMol & $f$-RAG & Gen.~GFN & Mol~GA & REINVENT \\
\midrule
albuterol\_similarity & $\mathbf{0.9778}$ & 0.937 & 0.977 & 0.949 & 0.896 & 0.882 \\
amlodipine\_mpo & $\mathbf{0.8105}$ & 0.810 & 0.749 & 0.761 & 0.688 & 0.635 \\
celecoxib\_rediscovery & $\mathbf{0.8747}$ & 0.826 & 0.778 & 0.802 & 0.567 & 0.713 \\
deco\_hop & $0.9279$ & \textbf{0.960} & 0.936 & 0.733 & 0.649 & 0.666 \\
drd2 & $\mathbf{0.9994}$ & 0.995 & 0.992 & 0.974 & 0.936 & 0.945 \\
fexofenadine\_mpo & $0.8720$ & \textbf{0.894} & 0.856 & 0.856 & 0.825 & 0.784 \\
gsk3b & $\mathbf{0.9930}$ & 0.986 & 0.969 & 0.881 & 0.843 & 0.865 \\
isomers\_c7h8n2o2 & $0.9511$ & 0.942 & 0.955 & \textbf{0.969} & 0.878 & 0.852 \\
isomers\_c9h10n2o2pf2cl & $0.8842$ & 0.833 & 0.850 & \textbf{0.897} & 0.865 & 0.642 \\
jnk3 & $0.8054$ & \textbf{0.906} & 0.904 & 0.764 & 0.702 & 0.783 \\
median1 & $\mathbf{0.4096}$ & 0.398 & 0.340 & 0.379 & 0.257 & 0.356 \\
median2 & $\mathbf{0.3897}$ & 0.359 & 0.323 & 0.294 & 0.301 & 0.276 \\
mestranol\_similarity & $\mathbf{0.9908}$ & 0.982 & 0.671 & 0.708 & 0.591 & 0.618 \\
osimertinib\_mpo & $\mathbf{0.9146}$ & 0.876 & 0.866 & 0.860 & 0.844 & 0.837 \\
perindopril\_mpo & $0.7126$ & \textbf{0.718} & 0.681 & 0.595 & 0.547 & 0.537 \\
qed & $\mathbf{0.9471}$ & 0.942 & 0.939 & 0.942 & 0.941 & 0.941 \\
ranolazine\_mpo & $\mathbf{0.8253}$ & 0.821 & 0.820 & 0.819 & 0.804 & 0.760 \\
scaffold\_hop & $\mathbf{0.6588}$ & 0.628 & 0.576 & 0.615 & 0.527 & 0.560 \\
sitagliptin\_mpo & $\mathbf{0.7054}$ & 0.584 & 0.601 & 0.634 & 0.582 & 0.021 \\
thiothixene\_rediscovery & $\mathbf{0.7315}$ & 0.692 & 0.584 & 0.583 & 0.519 & 0.534 \\
troglitazone\_redisc. & $0.8196$ & \textbf{0.867} & 0.448 & 0.511 & 0.427 & 0.441 \\
valsartan\_smarts & $\mathbf{0.9017}$ & 0.822 & 0.627 & 0.135 & 0.000 & 0.178 \\
zaleplon\_mpo & $0.5653$ & \textbf{0.584} & 0.486 & 0.552 & 0.519 & 0.358 \\
\midrule
Reported aggregate & \textbf{18.668} & 18.362 & 16.928 & 16.213 & 14.708 & 14.196 \\
\bottomrule
\end{tabular}
\end{table}

We evaluated five protein targets, three starting molecules per target and two molecular-similarity thresholds, giving 30 cells. A successful candidate must satisfy the Tanimoto-similarity, QED and SA constraints and improve the starting molecule's QuickVina2 docking score. At similarity thresholds of 0.4 and 0.6, STAMP returned improving candidates in 15/15 and 12/15 cells, respectively. GenMol succeeded in 14/15 and 12/15 cells, and Graph GA in 10/15 and 9/15 cells (Fig.~\ref{fig:optimization}c). Supplementary Note~4 gives the docking protocol and the complete per-cell comparison, including RetMol.

Across the 27 cells in which at least one method succeeded, STAMP achieved the best docking score in 23, GenMol in three and Graph GA in one. In the pairwise comparison with GenMol, STAMP gave the better outcome in 24 cells, either through a lower score or by returning an improving candidate when GenMol did not.

The atom-level action produced the 5ht1b/2 result at similarity threshold 0.6, extending the fragment search to a protonation-sensitive change.

\subsection{Compiled actions improve instruction-guided editing}

\begin{figure}[tbp]
\centering
\includegraphics[width=\linewidth,height=0.68\textheight,keepaspectratio]{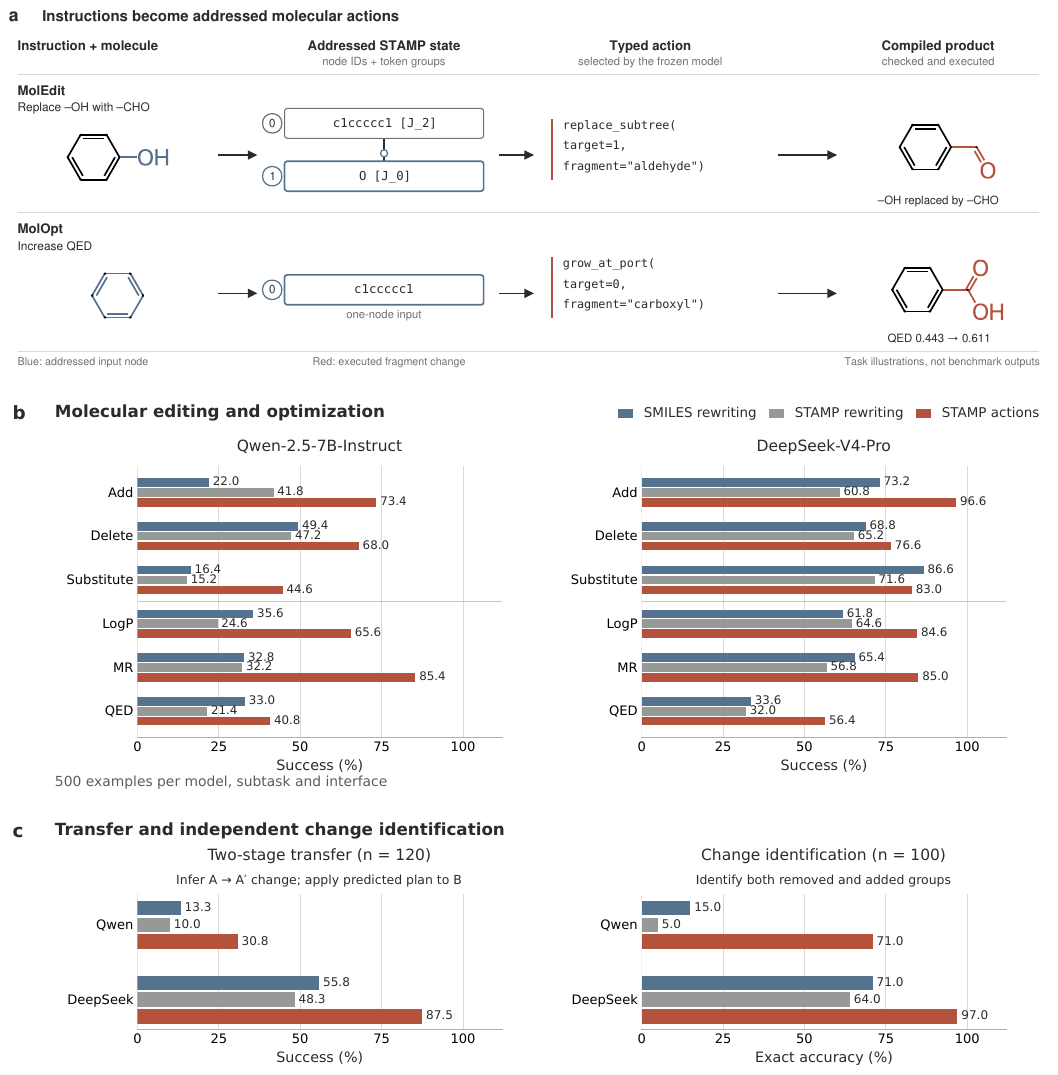}
\caption{\textbf{Compiled fragment actions support instruction-guided molecular editing.} \textbf{a}, An instruction is converted into an action on an indexed STAMP state and executed by the compiler. The MolEdit example specifies hydroxyl-to-aldehyde replacement; the MolOpt example specifies increased QED and carboxyl growth. Blue marks the addressed node and red the executed change. Molecules are task illustrations. \textbf{b}, Success on all 500 examples in each of six S\textsuperscript{2}-Bench subtasks for frozen Qwen-2.5-7B-Instruct and DeepSeek-V4-Pro, comparing STAMP subtree actions with complete-SMILES and complete-STAMP rewriting. Success tests group-count changes for MolEdit and property direction for MolOpt. \textbf{c}, Two-stage transfer on 120 examples per model and independent change identification on 100 pairs per model. Transfer predicts a removed/added plan from A/A', then edits B using that plan in a separate model call. Identification matches chemical information across representations, requires both group names to be correct and executes no edit. STAMP uses port-preserving single-node replacement for transfer and node views for identification. Atom-indexed controls and shared-label transfer results are in Supplementary Note~11.}
\label{fig:lm_interface}
\end{figure}

We tested frozen Qwen-2.5-7B-Instruct \cite{yang2024qwen25} and DeepSeek-V4-Pro \cite{deepseekai2026deepseekv4} on six S\textsuperscript{2}-Bench subtasks, using all 500 examples per subtask \cite{li2024s2bench}. MolEdit specifies a functional group to add, delete or substitute; MolOpt specifies a direction of change in LogP, molar refractivity or QED. The STAMP node interface exposes motifs, functional-group annotations and tree structure, then compiles a selected subtree action into a molecular edit (Fig.~\ref{fig:lm_interface}a). Complete-SMILES and complete-STAMP rewriting provide string-based controls. These conditions present different output tasks to the same frozen model. Rewriting requires a complete product sequence, including the retained molecular context. The action interface asks the model to choose a node, an operation and a replacement or growth fragment; the compiler reconstructs the product from that specification. The comparison therefore tests whether direct access to fragment targets improves execution of a molecular instruction.

Across the three MolEdit subtasks, mean success increased from 29.3\% with complete-SMILES rewriting to 62.0\% with STAMP actions for Qwen-2.5-7B-Instruct. DeepSeek-V4-Pro increased from 76.2\% to 85.4\% (Fig.~\ref{fig:lm_interface}b). The largest gains occurred in addition, reaching 73.4\% for Qwen and 96.6\% for DeepSeek, compared with 22.0\% and 73.2\% using SMILES. Gains were task-dependent: DeepSeek substitution reached 83.0\% with STAMP actions versus 86.6\% with SMILES rewriting.

On MolOpt, mean success increased from 33.8\% to 63.9\% for Qwen and from 53.6\% to 75.3\% for DeepSeek. QED-directed editing remained the most difficult STAMP subtask, with success rates of 40.8\% and 56.4\%, respectively. Complete-STAMP rewriting achieved lower family-level success than STAMP actions for both models, distinguishing the executable node interface from the serialized language alone. Success is determined by the requested group-count or property-direction change.

Transformation transfer used two independent stages: infer the change from A and A', then edit B using only that inferred plan. Across 120 examples per model, STAMP node replacement succeeded on 30.8\% for Qwen and 87.5\% for DeepSeek, compared with 13.3\% and 55.8\% using complete-SMILES rewriting (Fig.~\ref{fig:lm_interface}c and Supplementary Table~10).

In a separate 100-pair identification test with matched chemical information across representations, exact removed-and-added group accuracy was 71\% and 97\%, versus 15\% and 71\% with SMILES (Supplementary Table~11). Each molecular view was generated independently, so the model had to identify the change from the two structures. This result links the fragment-tree view to interpretation of a molecular transformation, while the transfer task tests the subsequent application of an inferred transformation to a new molecule.

Independent atom-indexed SMILES actions are reported in Supplementary Note~11 and Table~9. This comparison shows that the advantage depends on the model and action interface, with smaller differences for DeepSeek and reversals in some controls. Together, the editing, transfer and identification results show how an indexed fragment-tree state supports both interpreting a chemical change and turning that change into an executable action.

\FloatBarrier
\section{Discussion}

STAMP establishes a shared fragment-tree state for molecular learning and editing. t-SMILES also makes fragment topology explicit, serializing fragment SMILES through a breadth-first traversal of a full binary tree. Its TSSA, TSDY and TSID variants encode connections through shared atoms, unnumbered dummy atoms and labelled dummy atoms, respectively \cite{wu2024tsmiles}. STAMP separates canonical motif identity from attachment-site indices and uses depth-first traversal to place each descendant subtree in a contiguous sequence span. This lets a motif recur across attachment contexts while giving a selected fragment group explicit replacement boundaries and connection requirements. The same units and relations therefore support molecular learning, subtree search and compiled editing.

The common state connects tasks through their chemical units. In a learned encoder, a motif token acquires a contextual representation from the surrounding fragment tree. In search or instruction-guided editing, the corresponding node identifies a structural target, and its ports specify how a replacement joins the molecule. Molecular features and edit locations are thus expressed over the same decomposition. This enables separate generative models, property encoders and frozen language models to interact with a common molecular structure through task-specific objectives and operations. Addressability makes that shared structure usable at the point where a prediction becomes a proposed molecular change.

For learning, the shared state supports both assembling fragments into molecules and contextualizing them for property prediction. STAMP's quality-yield advantage in the controlled comparison extended to larger-scale generation and several fragment-constrained tasks; text-conditioned generation also improved validity and exact matching. Adapted encoders improved most comparisons with SMILES-BPE on MoleculeNet and MoleculeACE, while SVM--ECFP4 remained the strongest aggregate classical reference on MoleculeACE.

The frozen embeddings identify chemical relationships available before task adaptation. Similar motifs occupy neighbouring regions, attachment changes produce smaller shifts than motif substitutions, and recurring transformations produce consistent displacements across contexts. Whole-molecule distances also track structural change more closely than in the matched SMILES-BPE baseline. After matching target and structural distance, frozen-distance scores remained close to chance for activity-cliff discrimination. Together with the supervised gains, these results distinguish structural organization in the pretrained space from activity prediction after supervised adaptation.

Search and instruction-guided editing use the same structural relations to define executable changes. Subtree spans localize replacements, and attachment signatures specify how fragments join the retained structure. In STAMP-Edit, the model selects a change and a chemical executor constructs the product. Atom-indexed SMILES actions can also support effective editing, particularly with DeepSeek; the observed gains reflect the complete input and action interfaces. STAMP therefore offers a useful fragment-level organization for action, with benefits that depend on model capability, annotations and execution rules.

The task differences show where fragment-level operations were most effective. Generation gains were concentrated in linker design, motif extension and superstructure generation; scaffold-decoration quality was slightly lower than GenMol's. The largest editing gains over complete-SMILES rewriting occurred in addition, while QED-directed changes remained difficult and DeepSeek substitution showed no gain. Property-directed editing requires both executable operations and the selection of a useful chemical change. Atom-level actions complement fragment operations: optimization can change hydrogen count or formal charge, altering motif identity, and the node-port STAMP interface can replace a single motif while preserving its boundary connections. Fragment languages also differ in the relations they expose. Molcano's junction tags express face-to-face and substructure-level connections \cite{na2026molcano}; STAMP's ordered tree makes subtrees directly addressable. The choice of molecular units and relations therefore shapes the operations available for construction and editing.

These findings motivate evaluating molecular languages both as representations and as interfaces for action. Reconstruction, compactness and learned chemical organization describe what the language preserves and makes accessible; locality and executable transformations describe how it supports change. By combining these roles in a fragment-tree state, STAMP connects specialized molecular models, search algorithms and general-purpose language models through a common set of chemical units.

\section{Methods}

\subsection{STAMP representation, encoding and decoding}

\paragraph{State definition.}
STAMP uses the token families $\mathcal{V}=\mathcal{V}_{\mathrm{mot}}\cup\mathcal{V}_{\mathrm{J}}\cup\mathcal{V}_{\mathrm{B}}\cup\mathcal{V}_{\mathrm{ctrl}}$, where $\mathcal{V}_{\mathrm{mot}}$ contains motif tokens, $\mathcal{V}_{\mathrm{J}}=\{[\mathrm{J}_i]:i\geq0\}$ contains joint tokens, $\mathcal{V}_{\mathrm{B}}=\{[\mathrm{B}_n]:n\geq1\}$ contains branch tokens, and $i$ and $n$ are integers. Control and stereochemical symbols form $\mathcal{V}_{\mathrm{ctrl}}$. Each trained model uses a finite subset of these token families. Let $\mathcal{M}_{\mathrm{enc}}$ denote the molecular graphs accepted by the chosen fragmentation and encoding configuration. The encoder $\mathcal{E}:\mathcal{M}_{\mathrm{enc}}\to\mathcal{V}^{*}$ emits a token sequence, and the deterministic decoder $\mathcal{D}$ returns a molecular graph or reports failure. Round-trip evaluation measures the fraction of input molecules satisfying $\mathcal{D}(\mathcal{E}(m))=m$ under the molecular-identity criteria below.

\paragraph{Fragmentation.}
Default fragmentation cleaves every single bond outside a ring, following the fragSMILES protocol. Rings and fused-ring systems remain within fragments, and the cut bonds define the reduced inter-fragment graph. Each fragment is converted to an RDKit-canonical SMILES without dummy atoms or atom-map numbers \cite{rdkit}, so chemically identical fragments share a motif identity across attachment contexts. The codec also supports BRICS \cite{degen2008brics}, MMPA \cite{griffen2011mmp}, rotatable-bond and learned principal-subgraph fragmentation \cite{geng2023micam}.

\paragraph{Rooted fragment tree and traversal.}
The reduced graph is rooted at the fragment with the highest degree, with ties broken by heavy-atom count. Children are ranked by decreasing subtree size. The highest-ranked child forms the main continuation, and the remaining children form branches. Each branch is emitted in depth-first order before the main continuation, so every subtree occupies a contiguous span. Alternative roots and child orders provide sequence augmentation during training. Supplementary Note~1 and Supplementary Algorithm~1 specify tree construction.

\paragraph{Token grammar and encoding.}
Encoding emits one motif group per tree node. Each group contains a motif token followed by its joint tokens and associated stereochemical symbols. A joint token $[\mathrm{J}_i]$ refers to heavy atom $i$ in the canonical atom order of the preceding motif. All joints in the root group denote output ports. In each non-root group, the first joint denotes the input port connecting to the parent; remaining joints denote output ports for descendants. Output joints are listed in traversal order, with branch ports preceding the main-continuation port. Before each branch subtree, the encoder emits $[\mathrm{B}_n]$, where $n$ is the number of motif groups in that subtree, including groups in nested branches. Thus an eight-group subtree uses $[\mathrm{B}_8]$. Joint, branch and stereochemical tokens do not contribute to $n$; $n$ counts subtree groups rather than the number of children at a node. A model can use a branch symbol only if that symbol belongs to its tokenizer vocabulary.

\paragraph{Stack-machine decoding.}
The decoder processes the sequence from left to right while maintaining the partial molecular graph $G$, a stack $S$ of open ports and a stack of branch frames. It places the root motif in $G$ and pushes its output ports in reverse traversal order. For each subsequent motif group, it pops a parent port, connects it to the group's input joint and pushes the remaining output joints in reverse order. On encountering $[\mathrm{B}_n]$, it pops the next attachment port, saves the remaining port stack and starts a branch-local stack containing that port. The branch frame stores the saved stack and a counter initialized to $n$. Each decoded motif group decrements the counters of all active frames, including enclosing branches. Completed frames are closed from the innermost outwards, restoring their saved port stacks. Supplementary Algorithm~2 gives the procedure.

For malformed generated sequences, an out-of-range joint index is reduced modulo the motif atom count, and a saturated attachment site advances to the next atom with available valence. An overlong branch is truncated to the remaining groups. A motif encountered with an empty port stack starts a new component, and unresolved output ports at termination are capped with implicit hydrogens. Outputs that fail decoding, sanitization or the connected-molecule requirement are invalid in the generation benchmarks.

\paragraph{Motif vocabularies.}
Vocabulary coverage depends on the fragmentation scheme. The tree representation keeps ring and non-single bonds within motifs; irreducible fragments therefore require matching vocabulary entries. Atom-level fallback applies to fragments that can be reduced under these constraints. Vocabularies are constructed from the corresponding training corpora. The controlled ZINC-250K comparison \cite{sterling2015zinc15} uses 6,783 total STAMP tokens. The larger-scale and fragment-infill AO-GPT models use 12,000 retained motifs and atom-level fallback, giving 12,659 total motif, structural, fallback and control tokens. For the molecular encoder, PS-VAE principal-subgraph mining \cite{kong2022psvae} extracts 9,904 motifs from a 20,000-molecule training subset. Frequency profiling on the cleaned 9.69M-molecule corpus selects 4,666 motifs, giving 4,761 total tokenizer entries. Cumulative token-occurrence coverage is 99\%, and motif-only coverage is 96.85\% (Supplementary Note~5).

\subsection{Autoregressive generation of the STAMP state}

\paragraph{Controlled representation model.}
The eight-representation comparison uses a causal next-token Transformer \cite{vaswani2017transformer} with four layers, model dimension 256, eight attention heads, feed-forward dimension 1,024, maximum length 256 and dropout 0.1. Each model receives the same 100,000 training and 10,000 validation inputs for 10 epochs, using batch size 256, AdamW learning rate 3 \ensuremath{\times} 10\textsuperscript{-4}, weight decay 0.01 and training seed 42. Molcano has two training encoding failures; the other representations retain all inputs. The backbone is fixed; vocabulary-dependent embedding and output dimensions follow each representation (Supplementary Table~2).

\paragraph{External representation adapters.}
The t-SMILES baseline \cite{wu2024tsmiles} uses TSDY with MMPA\_DY fragmentation and random reconstruction. The Molcano baseline \cite{na2026molcano} uses the SSSR-plus-acyclic-residue converter without a predefined fragment library. Both codecs use a reversible SMILES lexer. TSDY decoding requires a valid fragment tree and attachment structure without repair or fallback. Molcano assembly requires valid, complete and consistent atom mapping and preservation of element counts. Accepted de novo products must sanitize as connected molecules. TSDY's unnumbered attachment sites allow alternative reconstructions, and the Molcano converter has incomplete stereochemistry handling during fragment merging. Round-trip evaluation reports valid decoding and stereochemistry-sensitive molecular identity separately (Supplementary Table~1).

\paragraph{Larger-scale and fragment-infill models.}
De novo generation and fragment infilling use separate checkpoints of an any-order autoregressive Transformer, AO-GPT \cite{xue2025aogpt}. The model has 12 layers, dimension 768, 12 attention heads, feed-forward dimension 3,072 and maximum sequence length 256, with adaptive layer-normalization conditioning on the target position. Attention is causal in the sampled prediction order. The de novo result uses the checkpoint at 300,000 training steps, with global batch size 1,024 and learning rate 3 \ensuremath{\times} 10\textsuperscript{-4}. The fragment model is initialized from separate pretraining and fine-tuned on 980,127 infill examples for three epochs, with batch size 256 and learning rate 10\textsuperscript{-4}. It supplies linker design, motif extension and superstructure generation. Scaffold decoration adds 10 epochs of fine-tuning on 294,160 examples, using batch size 256 and learning rate 5 \ensuremath{\times} 10\textsuperscript{-5}. Supplementary Table~5 summarizes the configurations.

\paragraph{AO-GPT objective.}
Let \(x\) denote the input sequence and \(y\) its shifted labels, with prompt and padding labels ignored. For a sampled position permutation \(\pi\), let \(z_{\theta,j}(x;\pi)\) be the output logits at permuted position \(j\), and let \(\mathcal{J}_{\pi}\) contain positions with non-ignored labels. The objective is

\begin{equation}
\mathcal{L}(\theta)=\mathbb{E}_{\pi}\!\left[-\frac{1}{|\mathcal{J}_{\pi}|}\sum_{j\in\mathcal{J}_{\pi}}\log\!\left(\operatorname{softmax} z_{\theta,j}(x;\pi)\right)_{y_{\pi(j)}}\right].
\label{eq:ao}
\end{equation}

The sampler uses random permutations for 90\% of examples and the identity order for the remainder, keeping the first and last non-ignored positions fixed. De novo training predicts the full sequence. Fragment-constrained examples remove group-aligned spans to form a retained-context prompt and completion; prompt positions are excluded from the loss. The model predicts individual tokens within these spans.

\paragraph{De novo sampling.}
The controlled and larger-scale de novo experiments use left-to-right decoding. The controlled comparison requests 10,000 samples per representation in a single run at temperature 1.0, top-k 50 and no nucleus truncation. The larger-scale AO-GPT model requests 1,024 samples for each of five seeds at temperature 0.7, top-p 0.95 and no top-k truncation.

\paragraph{Fragment-constrained sampling.}
Prompts specify retained motif groups and holes at the requested locations; generated spans are spliced into the holes before decoding. Linker design and scaffold morphing use the same formulation. Linker design, motif extension and scaffold decoration each use 10 query molecules and 100 requests per query and seed. Superstructure generation uses 25 requests per attachment-site prompt across 39 prompts, aggregated to the 10 original queries. All tasks use temperature 1.3 and no top-k truncation; top-p is 0.95 for scaffold decoration and 1.0 otherwise. Diversity-prefix penalties are 4 for linker design and motif extension, 2 for scaffold decoration and 3 for superstructure generation. Scaffold holes are completed sequentially. Metrics are averaged over queries and then summarized across five seeds. Fragment preservation and constraint-aware quality are defined under Generation metrics.

\subsection{STAMP for molecule--text construction}

\paragraph{STAMP-MolT5 architecture and pretraining.}
We replace the SMILES molecular modality in MolT5 \cite{edwards2022molt5} with single-view STAMP sequences. The encoder--decoder is initialized from pretrained T5.1.1-Base weights, with 12 encoder and 12 decoder layers, dimension 768, feed-forward dimension 2,048 and 12 attention heads. A 30,000-entry motif vocabulary is mined from ZINC by principal-subgraph merging. Adding 30,549 STAMP motif, structural and control tokens to the 32,100-token T5 vocabulary gives 62,649 entries. Input and output embeddings are tied, and the expanded model has 246.3M parameters.

Pretraining alternates equally between C4 text and 100M ZINC15 molecular sequences using the T5 span-denoising objective. Maximum source and target lengths are 192 and 96. Training uses global batch size 1,024 across four H200 GPUs, learning rate 1.4 \ensuremath{\times} 10\textsuperscript{-4}, 5,000 warm-up steps, linear decay and seed 42. Pretraining lasts 500,000 steps, and the final checkpoint initializes both downstream directions.

\paragraph{Molecule--text translation.}
Separate models are fine-tuned for molecule-to-text captioning and text-to-molecule generation on the official ChEBI-20 split \cite{edwards2021text2mol} used by MolT5, with 3,300 test pairs. Both directions use sequence-to-sequence cross-entropy, source and target lengths of 256, batch size 32, learning rate 5 \ensuremath{\times} 10\textsuperscript{-4}, 5\% warm-up, linear decay and seed 42. Training lasts at most 30 epochs and stops after six validation evaluations without improvement. Beam search uses width 5 and at most 256 new tokens.

Captioning follows the official MolT5 evaluation: SciBERT WordPiece tokenization for BLEU-2, BLEU-4 and METEOR, and Porter stemming for ROUGE-1, ROUGE-2 and ROUGE-L. For text-to-molecule generation, STAMP predictions are decoded before evaluation. Validity is the fraction of all predictions yielding an RDKit-parseable molecule. Exact match compares RDKit-canonical isomeric SMILES over all examples. Character-level BLEU-4 and Levenshtein distance compare decoded and canonical reference strings. MACCS, RDKit and 2,048-bit Morgan fingerprints with radius 2 \cite{rogers2010ecfp} are compared by Tanimoto similarity \cite{bajusz2015tanimoto} over parseable prediction--reference pairs. Outputs outside the STAMP grammar are invalid. Published T5-Base \cite{raffel2020t5} and MolT5 results provide the reference values.

\subsection{Optimization by compatible subtree exchange}

\paragraph{TreeCrossover.}
Each subtree is addressable as a contiguous depth-first span (Algorithm~\ref{alg:crossover} and Fig.~\ref{fig:optimization}a). TreeCrossover grafts a peripheral donor subtree onto a recipient. A subtree \(u\) has port signature \(\sigma(u)=(b_u,p_u)\), where \(b_u\) is its incoming bond order and \(p_u\) is the number of open ports on its root motif. A donor \(b\) can replace recipient subtree \(a\) when \(\sigma(a)=\sigma(b)\) and \(|z(a)-z(b)|\leq\delta\), where \(z\) is the heavy-atom count and the default tolerance \(\delta\) is 4. Offspring are decoded and sanitized; invalid or unchanged proposals are rejected. Lead optimization freezes the starting scaffold and modifies peripheral subtrees.

Optimization supports compatible subtree exchange and atom-level changes throughout the search. Atom-level actions select an atom in the decoded molecule and change its hydrogen count or formal charge, followed by sanitization and evaluation. These changes alter the identity of the affected motif.

\paragraph{Controlled splice-operator comparison.}
We sampled 5,000 connected ZINC-250K molecules and parsed them with the default exocyclic-single-bond decomposition. With seed 0, we drew 1,000 independent recipient--donor trials for each of four operators: compatible STAMP subtree replacement, arbitrary STAMP token-span replacement, SAFE fragment-span replacement and SMILES character-substring replacement. Candidate spans were capped at 16 tokens or characters, and the compatible subtree operator used \ensuremath{\delta} = 4. Strict validity requires decoding without fallback followed by RDKit sanitization. Figure~\ref{fig:optimization}b reports strict-valid offspring and offspring that are both strict-valid and retain the recipient Bemis--Murcko scaffold \cite{bemis1996scaffold}, each divided by all 1,000 trials. Conditional retention uses scaffold-evaluable strict-valid products as its denominator.

\begin{algorithm}[t]
\caption{STAMP subtree crossover (\textsc{TreeCrossover})}\label{alg:crossover}
\begin{algorithmic}[1]
\Require parent SMILES $s_A, s_B$; encoder $\mathcal{E}$; size tolerance $\delta$; retries $T$
\Ensure offspring SMILES, or $\varnothing$
\State $P_A \gets \textsc{Parse}(\mathcal{E}(s_A))$; $P_B \gets \textsc{Parse}(\mathcal{E}(s_B))$
\State $S_A \gets \textsc{EditSpans}(P_A)$; $S_B \gets \textsc{EditSpans}(P_B)$ \Comment{subtrees, excluding root}
\For{$t \gets 1$ \textbf{to} $T$}
 \State $b \gets \textsc{Sample}(S_B)$; $\sigma_B \gets \textsc{PortSig}(P_B, b)$; $z_B \gets \textsc{HeavyAtoms}(P_B, b)$
 \State $H \gets \{ a \in S_A : \textsc{PortSig}(P_A,a)=\sigma_B \wedge |\textsc{HeavyAtoms}(P_A,a)-z_B|\le\delta \}$
 \If{$H = \emptyset$} \textbf{continue} \EndIf
 \State $a \gets \textsc{Sample}(H)$; $d \gets \textsc{StripLeadingBranch}(P_B.\text{tokens}[b])$
 \State $\tau \gets \textsc{SpliceDonor}(P_A.\text{tokens}, a, d)$; $s' \gets \textsc{Decode}(\tau)$
 \If{$s' \neq \varnothing \wedge s' \neq s_A$} \Return $s'$ \EndIf
\EndFor
\State \Return $\varnothing$
\end{algorithmic}
\end{algorithm}

\paragraph{Optimization loop and prescreening.}
The online loop starts from an objective-scored population drawn from the training distribution. At each generation, parents are sampled, TreeCrossover proposes an offspring and the oracle evaluates each unique decoded molecule once. The top-K molecules form the next population. High-reward molecules contribute compatible subtrees to a signature-keyed donor pool, with a cap of 50 per signature and 5,000 overall, retained by descending reward.

STAMP uses the same target-aware prescreening protocol as GenMol: approximately 250,000 ZINC-250K molecules are scored offline to build the initialization bank. PMO then uses a 10,000-call online oracle budget. A common hyperparameter configuration was used for all 23 objectives. Table~\ref{tab:pmo_main} reports STAMP means over three seeds. Supplementary Algorithm~3 gives the online loop.

\subsection{Representation learning and molecular property prediction}

\paragraph{Pretraining.}
STAMP uses the ChemBERTa-2 10M encoder configuration: six layers, dimension 768, 12 attention heads, feed-forward dimension 1,536 and sequence length 128. Pretraining uses the cleaned 9.69M-molecule PubChem-10M corpus \cite{ahmad2022chemberta2,kim2023pubchem} and the encoder vocabulary described above. STAMP-MTR and the released SMILES-BPE comparator use matched model configurations and pretraining objectives.

MLM selects a random subset \(\mathcal{M}\) comprising 15\% of tokens. Of these selected tokens, 80\% are masked, 10\% are replaced at random and 10\% are unchanged. The objective is

\begin{equation}
\mathcal{L}_{\mathrm{MLM}}(\theta)=\mathbb{E}_{\mathcal{M}}\sum_{i\in\mathcal{M}}-\log p_\theta\!\big(x^{(i)}\mid x_{\setminus\mathcal{M}}\big)
\label{eq:mlm}
\end{equation}

MLM pretraining uses 10 epochs, batch size 512 and learning rate 5 \ensuremath{\times} 10\textsuperscript{-4}.

MTR maps the pooled encoder representation \(g_\theta(x)\in\mathbb{R}^{768}\) to descriptor predictions \(\hat{\mathbf z}=h_\phi(g_\theta(x))\) using a prediction head \(h_\phi:\mathbb{R}^{768}\rightarrow\mathbb{R}^{208}\). The targets are 208 RDKit descriptors, with Ipc and Kappa3 excluded because of overflow. Descriptors are z-scored across the corpus to \(\tilde{\mathbf z}\), and the objective is \(\mathcal{L}_{\mathrm{MTR}}(\theta,\phi)=\tfrac{1}{208}\lVert\hat{\mathbf z}-\tilde{\mathbf z}\rVert_2^2\). MTR pretraining uses five epochs, batch size 512 and learning rate 3 \ensuremath{\times} 10\textsuperscript{-4}. Supplementary Table~5 summarizes the training configurations, validation metrics and training times.

\paragraph{Motif-neighbour analysis.}
For each motif identity, we average final-layer STAMP-MTR hidden states across molecular contexts and retrieve neighbours by cosine distance. Chemical similarity is ECFP4 Tanimoto similarity between motif graphs \cite{rogers2010ecfp}. We report mean similarity among the first k neighbours for k \ensuremath{\in} \{1, 3, 5, 10, 20\}, using random motif pairs as the reference. Illustrative queries span recurring ring and functional-group classes.

\paragraph{Frozen representation analysis.}
Molecular embeddings from STAMP-MTR and the released ChemBERTa-2 10M SMILES-BPE encoder \cite{ahmad2022chemberta2} are \ensuremath{\ell_2}-normalized before cosine-distance calculation. Comparisons use only pairs with finite embeddings for both molecules from both encoders. The controlled attachment panel combines ten aromatic parent structures with nine probe motifs. Attachment-position pairs vary the substituted parent atom while fixing the parent and probe; motif-identity pairs vary the probe while fixing the parent and substituted atom. Canonical duplicate pairs are removed. Unrelated pairs are sampled across different parents without overlap with the controlled categories. The resulting panel contains 207 attachment-position, 1,080 motif-identity and 4,000 unrelated pairs.

For transformation directions, matched molecular pairs are oriented by lexicographic order of the variable fragments and collapsed to one observation per canonical transformation and unique context. Transformations observed in at least six contexts are retained, with at most 100 contexts per transformation. The same transformations enter both encoder analyses. Each displacement is normalized and compared by cosine similarity with the centroid from the remaining contexts of that transformation. The null samples 5,000 pairs of displacements from different transformations, using identical pair indices for both encoders. Whole-molecule structural alignment is the Spearman correlation between embedding cosine distance and ECFP4 Tanimoto distance on MoleculeACE matched pairs with valid embeddings from both encoders. The structure-matched activity-cliff analysis and RePRA protocol are specified in Supplementary Notes~9 and~10.

\paragraph{Fine-tuning.}
The standard fine-tuning configuration uses batch size 32, learning rate 2 \ensuremath{\times} 10\textsuperscript{-5} and 20 epochs. MoleculeNet uses the scaffold splits from the ChemBERTa-2 evaluation; MoleculeACE uses its supplied train/test assignments. Classification heads are trained with cross-entropy and evaluated by ROC-AUC; regression is evaluated by RMSE. MoleculeNet compares fine-tuned STAMP encoders with published ChemBERTa-2 10M results under the corresponding MLM or MTR objective \cite{ahmad2022chemberta2}. MoleculeACE fine-tunes STAMP-MTR and the released SMILES-BPE encoder under the same downstream protocol and on identical molecule sets. Between-encoder RMSE differences across the 30 targets are assessed with paired Wilcoxon signed-rank tests.

\subsection{Shared evaluation and implementation}

\paragraph{Datasets and splits.}
The representation and controlled generation studies use ZINC-250K. Round-trip evaluation covers a common 20,000-molecule sample from the training split (seed 42) with SMILES, SELFIES \cite{krenn2020selfies}, Group-SELFIES \cite{groupselfies2023}, SAFE \cite{noutahi2024safe}, fragSMILES, t-SMILES \cite{wu2024tsmiles}, Molcano \cite{na2026molcano} and STAMP. Each encoded sequence is decoded and compared by canonical isomeric SMILES, with InChIKey equality as a fallback for equivalent encodings. Encoding, valid-decoding and exact-identity rates use all 20,000 sampled molecules as their denominator. STAMP encoded and exactly reconstructed every molecule in this sample (100\%). We also report token lengths and observed vocabulary sizes. The ChEMBL edit-locality analysis compares six representations (Supplementary Note~8). The ZINC20-mined corpus for larger-scale AO-GPT pretraining retained 425,457,500 training records and 51,000,653 validation records (476,458,153 records in total) \cite{lee2024genmol,irwin2020zinc20}. Validation used 20,000 records sampled from the retained validation split.

\paragraph{Baselines.}
SAFE-GPT and GenMol generation results are taken from GenMol \cite{lee2024genmol}. PMO baseline values are from GenMol, f-RAG and PMO \cite{lee2024genmol,gao2022pmo,frag2024}. Additional prescreened methods report aggregate AUC-Top10 scores of $18.993\pm0.219$ for InVirtuoGen and 19.270 for Graph-GRPO \cite{kaech2026invirtuogen,zhu2026graphgrpo}. Classical and deep-learning MoleculeACE baselines are from the benchmark authors \cite{vantilborg2022moleculeace}.

\paragraph{Generation metrics.}
Let \(R\) be the number of requests, \(V\) the number of valid decoded outputs and \(\mathcal{S}\) the set of \(N\) unique valid molecules after canonicalization. Validity is \(V/R\). Uniqueness is \(N/V\) in the controlled and fragment-constrained experiments and \(N/R\) for the larger-scale STAMP result. Novelty is the fraction of \(\mathcal{S}\) absent from the training corpus. Internal diversity uses Tanimoto similarity \(T(m_i,m_j)\) between radius-2, 2,048-bit Morgan fingerprints. Quality yield counts unique valid molecules satisfying both thresholds per request:

\begin{equation}
\begin{aligned}
\mathrm{Div}&=1-\frac{2}{N(N-1)}\sum_{i<j}T(m_i,m_j),\\
\mathrm{Q}&=\frac{1}{R}\sum_{m\in\mathcal{S}}\mathbb{1}\!\left[\mathrm{QED}(m)\!\ge\!0.6\,\wedge\,\mathrm{SA}(m)\!\le\!4\right].
\end{aligned}
\label{eq:metrics}
\end{equation}

Diversity is estimated using at most 5,000 unique molecule pairs. Fragment-constrained quality additionally requires each counted molecule to preserve the supplied constraints. Preservation rate is the fraction of unique valid molecules satisfying those constraints. Distance from the seed is the mean Tanimoto distance using radius-2, 1,024-bit Morgan fingerprints (Supplementary Fig.~1).

\paragraph{Optimization metrics.}
PMO reports AUC-Top10: the area under the running top-10 mean-reward curve over 10,000 online oracle calls, normalized to \([0,1]\). The lead-optimization benchmark contains five targets, three starting molecules per target and two similarity thresholds. Each cell reports the best QuickVina2 score among improving, constraint-satisfying molecules, averaged over three optimization seeds \cite{alhossary2015quickvina,trott2010vina}. Lower scores are better. The docking budget is 1,000 calls, with similarity, QED and SA constraints applied before docking. Figure~\ref{fig:optimization}c shows per-cell results and coverage; Supplementary Table~4 gives the full four-method comparison.

\paragraph{Implementation.}
The codec, vocabulary mining and evaluations use Python, RDKit and PyTorch. RDKit handles molecular input/output, canonicalization, fragmentation, sanitization, descriptors and fingerprints. Models are trained on H200 GPUs.

\subsection{Executing STAMP actions with frozen language models}

\paragraph{Models and evaluation design.}
Qwen-2.5-7B-Instruct \cite{yang2024qwen25} and DeepSeek-V4-Pro \cite{deepseekai2026deepseekv4} were accessed through OpenRouter using model IDs \path{qwen/qwen-2.5-7b-instruct} and \path{deepseek/deepseek-v4-pro} with temperature 0, a maximum of 8,192 output tokens and reasoning requested as disabled. Model weights remained frozen. Each condition used one complete evaluation run. All six S\textsuperscript{2}-Bench CSV files contributed their full 500 rows in the original order \cite{li2024s2bench}. Models and interfaces shared the same task snapshots. Each MolEdit or MolOpt response allowed one additional call only for parsing or execution failure, without task-correctness or property feedback.

\paragraph{STAMP and string interfaces.}
The primary STAMP interface presents canonical SMILES and a motif-node table with roles and functional-group labels from a fixed SMARTS library. Labels cover all matching input groups and are not restricted to the requested target. The model emits \texttt{delete\_subtree}, \texttt{replace\_subtree} or \texttt{grow\_at\_port}. Deletion and replacement affect the selected node and its descendants; root deletion and replacement are disallowed. Replacement requires one external boundary. Growth selects a connection site within the chosen node using fixed hydrogen-availability rules. The parser maps recognized group names to fixed fragment structures and resolves fragment attachment atoms by deterministic rules. Supplementary Note~11 specifies these rules, the independent SMILES atom-action baseline and shared-label transfer controls.

Subtree selections are mapped to atom sets and passed to the same RDKit graph-editing implementation used by the atom-indexed baseline. Thus both action interfaces have chemical executors. Complete-SMILES and complete-STAMP conditions instead generate the entire product string. For MolEdit and MolOpt, raw STAMP input also includes reference SMILES. The codec retains its standard decoding tolerance.

\paragraph{Task scoring.}
MolEdit success requires the named group's count to change by exactly +1 for addition or -1 for deletion; substitution requires both changes. MolOpt success requires a strict change in the requested direction in RDKit LogP, molar refractivity or QED, without a minimum magnitude or similarity threshold. All 500 cases remain in each denominator. S\textsuperscript{2}-Bench validity counts RDKit-parsable products; complete-SMILES outputs may therefore be multicomponent. The action and STAMP-decoding paths additionally enforce nonempty, connected products. Similarity is mean Morgan-fingerprint Tanimoto similarity among valid products, using radius 2 and 2,048 bits.

\paragraph{Two-stage transformation transfer.}
We constructed 120 A/A'/B examples from the substitution molecule pool and used the same examples for both models and all interfaces. First, the model sees A and A' and predicts removed and added group names. Valid but incorrect plans proceed unchanged; invalid plans terminate the example. A separate second-stage conversation receives only B and that condition's predicted plan. STAMP replaces one selected motif, retaining its neighbours and every original labelled boundary port; the model supplies the complete replacement fragment. Root motifs may be edited. Each stage uses one model call. Success requires a valid, connected product with the true removed-group count reduced by one and added-group count increased by one. Both success and validity use all 120 examples as the denominator.

\paragraph{Independent change identification.}
A separate 100-pair dataset requires one prediction of the removed and added groups, without editing or repair. Each molecule is encoded independently, with no atom correspondence, node alignment or preselected change location. Chemical information is matched across the STAMP node, atom-indexed SMILES and complete-string views. Exact accuracy requires both names to be correct. Supplementary Note~11 gives construction details, stage-level diagnostics and controls with shared atom-level labels.

\section{Data availability}

The data supporting this study are available at \url{https://huggingface.co/datasets/bi0s/STAMP_V1-data}. The study draws on publicly available molecular collections and benchmarks. ZINC-250K, ZINC15 and ZINC20 are used for representation analysis and molecular generation, and PubChem-10M is used for molecular-encoder pretraining. ChEMBL provides matched molecular pairs for edit-locality analysis. Molecular property prediction and activity-cliff evaluation use MoleculeNet and MoleculeACE. Molecule--text modelling uses C4 text and ZINC15 molecules for pretraining and ChEBI-20 for downstream training and evaluation. Language-model-directed editing and optimization use the MolEdit and MolOpt tasks from S\textsuperscript{2}-Bench.

The transfer dataset was constructed from input molecules in the S\textsuperscript{2}-TOMG-Bench substitution subset (\path{MolEdit_SubComponent.csv}). The identification dataset was constructed separately from a merged and deduplicated pool of input molecules from the addition, deletion, substitution and LogP subsets (\path{MolEdit_AddComponent.csv}, \path{MolEdit_DelComponent.csv}, \path{MolEdit_SubComponent.csv} and \path{MolOpt_LogP.csv}). The construction procedures for both datasets are described in Supplementary Note~11.

Dataset selection, preprocessing, train--validation--test partitions and evaluation subsets are described in Methods and Supplementary Information. The optimization objectives and lead-optimization benchmark are specified in the corresponding Methods sections. Published baseline results are attributed to their original studies in Methods and the relevant tables.

\section{Code availability}

The source code for the STAMP framework is available at \url{https://github.com/YuX-Ren/STAMP_V1}. The framework comprises STAMP encoding and deterministic decoding, motif-vocabulary construction, any-order autoregressive generation and fragment infilling, TreeCrossover optimization, molecular representation learning, molecule--text modelling and the STAMP-Edit interface for language-model-directed molecular editing. Molecular operations use RDKit, and neural models are implemented in PyTorch. Model architectures, training and fine-tuning objectives, hyperparameters, sampling settings and evaluation protocols are specified in Methods and Supplementary Information.

\section{Author contributions}

Y.R., J.L. and Y.B. developed the research concept. Y.R. designed the methods, algorithms and experiments and implemented the software. Y.R. and J.L. performed the experiments and analysed the results. Y.R. wrote the initial draft of the manuscript, and all authors contributed to manuscript writing and revision.

\end{document}